\documentclass[twocolumn,twoside,slac_two]{revtex4}
\usepackage{graphicx}
\usepackage{fancyhdr}
 \newcommand{\beq}{\begin{equation}}
 \newcommand{\eeq}{\end{equation}}
 \newcommand{\bqa}{\begin{eqnarray}}
 \newcommand{\eqa}{\end{eqnarray}}

\def\ket#1{\left| #1\right\rangle}

\newcommand{\Kz}{\mbox{$K^{0}$}}
\newcommand{\Kzbar}{\mbox{$\overline{K^{0}}$}}

\newcommand{\KL}{\mbox{$K_{L}$}}
\newcommand{\KS}{\mbox{$K_{S}$}}

\newcommand{\tauS}{\mbox{$\tau_{S}$}}

\newcommand{\delm}{\mbox{$\Delta m$}}

\def\epe{\epsilon'\!/\epsilon}

\newcommand{\epsrat}{\mbox{$\epsilon^{\prime}\!/\epsilon$}}
\newcommand{\reepoe}{\mbox{$Re(\epsrat)$}}
\newcommand{\imepoe}{\mbox{$Im(\epsrat)$}}
\newcommand{\ktev}{\mbox{KTeV}}
\newcommand{\etapm}{\mbox{$\eta_{+-}$}}
\newcommand{\etazz}{\mbox{$\eta_{00}$}}
\newcommand{\phipm}{\mbox{$\phi_{+-}$}}
\newcommand{\phizz}{\mbox{$\phi_{00}$}}
\newcommand{\phisw}{\mbox{$\phi_{SW}$}}
\newcommand{\phiep}{\mbox{$\phi_{\epsilon}$}}
\newcommand{\delphi}{\mbox{$\Delta \phi$}}

\def\kchrg{K\to\pi^+\pi^-}
\def\kneut{K\to\pi^0\pi^0}

\def\kpi0{K_{L}\to 3\pi^0}
\def\ke3{K_{L}\to\pi^{\pm}e^{\mp}\nu}
\def\km3{K_{L}\to\pi^{\pm}\mu^{\mp}\nu}
\def\k2pi{K_{L} \to \pi^+\pi^-}

\newcommand{\Kpp}{\mbox{$K\rightarrow\pi\pi$}}
\newcommand{\Kpm}{\mbox{$K\rightarrow\pi^{+}\pi^{-}$}}
\newcommand{\Kzz}{\mbox{$K\rightarrow\pi^{0}\pi^{0}$}}

\newcommand{\KLpp}{\mbox{$K_{L}\rightarrow\pi\pi$}}
\newcommand{\KSpp}{\mbox{$K_{S}\rightarrow\pi\pi$}}
\newcommand{\KLpm}{\mbox{$K_{L}\rightarrow\pi^{+}\pi^{-}$}}
\newcommand{\KSpm}{\mbox{$K_{S}\rightarrow\pi^{+}\pi^{-}$}}
\newcommand{\KLzz}{\mbox{$K_{L}\rightarrow\pi^{0}\pi^{0}$}}
\newcommand{\KSzz}{\mbox{$K_{S}\rightarrow\pi^{0}\pi^{0}$}}

\newcommand{\Kzzz}{\mbox{$K_L \rightarrow \pi^{0}\pi^{0}\pi^{0}$}}

\newcommand{\ppc}{\mbox{$\pi^{+}\pi^{-}$}}

\newcommand{\ppn}{\mbox{$\pi^{0}\pi^{0}$}}

\newcommand{\pzpz}{\pi^{0}\pi^{0}}

\newcommand{\keth}{\mbox{$\pi^{\pm} e^{ \mp}\nu_e$}}

\newcommand{\zzz}{\mbox{$\pi^{0}\pi^{0}\pi^{0}$}}

\newcommand{\eu}{ \times 10^{-4}}

\newcommand{\degs}{^{\circ}}
\newcommand{\delmunits}{\mbox{$\times 10^{6}~\hbar {\rm s}^{-1}$}}
\newcommand{\tausunits}{\mbox{$\times 10^{-12}~{\rm s}$}}

\def\dmswval{5269.9}
\def\tsswval{89.623}

\def\dmswerr{  12.3}
\def\tsswerr{ 0.047}

\def\dphcptval{ 0.30}

\def\dphcpterr{ 0.35}

\def\phpmval{ 43.76}
\def\phpmerr{  0.64}
\def\phzzval{ 44.06}
\def\phzzerr{  0.68}

\def\dphswval{  0.40}
\def\dphswerr{  0.56}

\begin{document}

\title{The Final Measurement of $\epe$ from KTeV}

%

\author{E. Worcester}
\affiliation{University of Chicago, Chicago, Illinois}

\begin{abstract}
We present precise measurements of CP and CPT symmetry
based on the full dataset of $\Kpp$ decays collected by the KTeV experiment 
at FNAL.
We measure the direct CP violation parameter$\reepoe$ = (19.2 $\pm 2.1)\eu$.
We find the $K_L$-$K_S$ mass difference $\delm$~=~(5265 $\pm$ 10)$\delmunits$
and the $K_S$ lifetime $\tauS$ = (89.62 $\pm$ 0.05)$\tausunits$.  We test CPT
symmetry by finding the phase of the indirect CP violation parameter 
$\epsilon$, $\phiep$~=~(44.09 $\pm$ 1.00)$\degs$, and the difference of the relative 
phases between the CP violating and CP conserving decay amplitudes for 
$\Kpm$ ($\phipm$) and for $\Kzz$ ($\phizz$), $\delphi$ = (0.29 $\pm$ 0.31)$\degs$.
These results are consistent with other experimental results and
with CPT symmetry.  

\end{abstract}

\maketitle

\thispagestyle{fancy}


\section{Introduction}
Violation of CP symmetry occurs in the neutral kaon system in
two different ways.  The dominant effect is the
result of an asymmetry in the mixing of $\Kz$ and $\Kzbar$ such
that $\KL$ and $\KS$ are not CP eigenstates.  This effect is
parameterized by $\epsilon$ and is called indirect CP violation.
The other effect, called direct CP violation, occurs in the
$\Kpp$ decay process and is parameterized by $\epsilon'$.
Direct CP violation affects the decay rates of $\Kpm$ and $\Kzz$
differently, so it is possible to measure the level of direct
CP violation by comparing $\etapm$ and $\etazz$:
\begin{equation}
 \begin{array}{ccccc}
\etapm & = &\frac{A(\KLpm)}{A(\KSpm)}\ & = & \epsilon + \epsilon'
\\
\\
\etazz & = &\frac{A(\KLzz)}{A(\KSzz)}\ & = & \epsilon - 2\epsilon'
\\
\\
\reepoe & \approx & \frac{1}{6}\ ( |\frac{\etapm}{\etazz}\ |^2 - 1).& &
\\
 \end{array}
\label{eq:reepoe}
\end{equation}
Measurements of $\pi\pi$ phase shifts~\cite{ochs} show that,
in the absence of CPT violation, the phase of $\epsilon'$ is
approximately equal to that of $\epsilon$. Therefore, $\reepoe $ is a
measure of direct CP violation and $\imepoe $ is a measure of CPT
violation.

For small $|\epsilon'/\epsilon|$, $\imepoe$ is related
to the phases of $\etapm$ and $\etazz$ by
\begin{equation}
\begin{array}{lcl}
  \phipm &\approx &\phiep  + \imepoe  \\
  \phizz &\approx &\phiep  - 2 \imepoe  \\
  \delphi &\equiv &\phizz  - \phipm   \approx -3  \imepoe~.
\end{array}
  \label{eq:delphimpe}
\end{equation}

Experimental results have established that $\reepoe$ is non-zero
\cite{prl:731,pl:na31,prl:pss,na48:reepoe}.  In 2003, KTeV
reported $\reepoe = (20.7 \pm 2.8)\eu$ based on data from 1996 and 
1997\cite{prd03}.
We now report the 
final measurement of $\reepoe$ from KTeV.
The measurement is based on 85 million reconstructed \Kpp\ decays
collected in 1996 1997, and 1999.
This full sample is two 
times larger than, and contains, the sample on which the previous
results are based.
We also present measurements of
the kaon parameters $\delm$ and $\tauS$,
and tests of CPT symmetry based on measurements of
$\delphi$ and $\phipm -\phisw$.

For these results we have made significant improvements to the data 
analysis and the Monte Carlo simulation.
The full dataset, including those data used in the previous analysis,
has been reanalyzed using the improved reconstruction and simulation.
These results supersede the previously published results
from KTeV\cite{prd03}.
In this presentation, we will focus primarily on improvements to the
neutral mode analysis which have reduced the systematic uncertainty
in $\reepoe$
relative to the previous KTeV result.

\section{The KTeV Experiment}
The measurement of \reepoe\ requires a source of $K_L$ and $K_S$
decays, and a detector to reconstruct the charged ($\ppc$)
and neutral ($\ppn$) final states.
The strategy of the \ktev\ experiment is to produce two identical
$K_L$ beams, and then to pass one of the beams through a
``regenerator.''
The beam that passes through the regenerator is called the
regenerator beam,
and the other beam is called the vacuum beam.
The regenerator creates a coherent
$\ket{K_L}+\rho\ket{K_S}$ state,
where $\rho$, the regeneration amplitude, is a physical property
of the regenerator.  The regenerator is designed
such that most of the \Kpp\ decays
downstream of the regenerator are from the $K_S$ component.
The charged spectrometer is the 
primary detector for reconstructing $\Kpm$ decays and the 
pure Cesium Iodide (CsI) calorimeter 
is used to reconstruct the four photons from $\Kzz$ decays.  
A Monte Carlo simulation is used to correct for the acceptance
difference between \Kpp\ decays in the two beams,
which results from the very different $K_L$ and $K_S$ lifetimes.
The measured quantities are the vacuum-to-regenerator
``single ratios''  for \Kpm\ and \Kzz\ decay rates.
These single ratios are proportional to
$|\etapm/\rho|^2$ and $|\etazz/\rho|^2$,
and the ratio of these two quantities gives $\reepoe$
via Eq.~\ref{eq:reepoe}.

\subsection{The KTeV Detector}
\begin{figure}
\begin{centering}
\includegraphics[width=80mm]{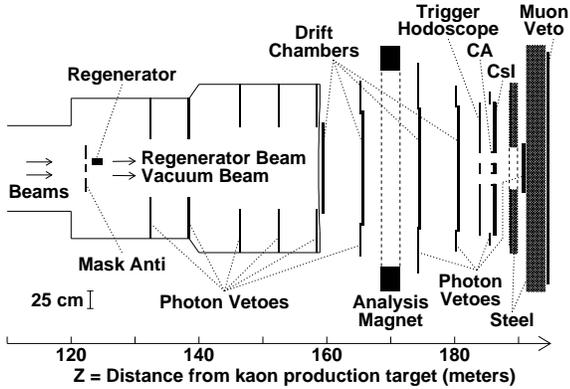}
\caption{The KTeV Detector}
\label{fig:det}
\end{centering}
\end{figure}

The KTeV detector (Figure \ref{fig:det}) consists of a charged spectrometer 
to reconstruct
$\Kpm$ decays, a pure CsI electromagnetic calorimeter to reconstruct
$\Kzz$ decays, a veto system to reduce background, and a three-level trigger to
select events.  Two virtually identical neutral kaon beams are incident
on the detector; a movable active regenerator is placed in
one of these beams to provide a coherent mixture of $\KL$ and $\KS$.
In this manner, we collect $\KLpp$ and $\KSpp$ decays simultaneously
so that many systematic effects cancel in the ratios used to calculate
\reepoe.  

The KTeV spectrometer consists of four drift chambers and a
large dipole magnet.  It measures the momenta of charged particles with
an average resolution of ~0.4\%.  The $\Kpm$ reconstruction achieves a 
$z$-vertex 
resolution of 5-30 cm and a mass resolution of 1.5 MeV/$c^{2}$.
The CsI calorimeter measures the energies and positions of photons from
the electromagnetic decay of the neutral pions in $\Kzz$ decays.  It has
an average energy resolution of 0.6\%.  The reconstructed decay vertex 
of the neutral
pion is directly related to the energies of the photons:
\begin{equation}
\label{eq:zcalc}
Z_{\pi^0} = Z_{CsI} - \frac{r_{12}\sqrt{E_{1}E_{2}}}{m_{\pi^0}}.\
\end{equation}
The $\Kzz$ reconstruction achieves a $z$-vertex resolution of 20-30 cm
and a mass resolution of 1.5 MeV/$c^{2}$.

\subsection{Monte Carlo Simulation}
KTeV uses a Monte Carlo (MC) simulation to calculate the detector
acceptance and to model background to the signal modes.  
The very different $\KL$ and $\KS$ lifetimes lead to very different 
$z$-vertex distributions in the 
vacuum and regenerator beams.  We determine the 
detector acceptance as a function of kaon decay vertex and energy including 
the effects of geometry, detector response, and resolutions.  
To help verify the accuracy of the MC simulation, we collect 
and study decay modes with approximately ten times higher statistics than 
the $\Kpp$ signal samples, such as $\ke3$ and \Kzzz.

Many improvements have been made to the MC simulation since 
the previous result was published in 2003\cite{prd03}.  We have improved
the simulation of electromagnetic showers
to include the effects of incident particle angles
and to simulate the effects of wrapping and shims in
the CsI calorimeter.
We have improved the tracing of charged particles through the detector
with more complete treatments of ionization energy loss, Bremsstrahlung,
delta rays, and hadronic interactions in the drift chambers.  
We have also have updated a number of parameters that go into the 
kaon propogation and decay calculations.

\begin{figure}
\begin{center}
\includegraphics[width=80mm]{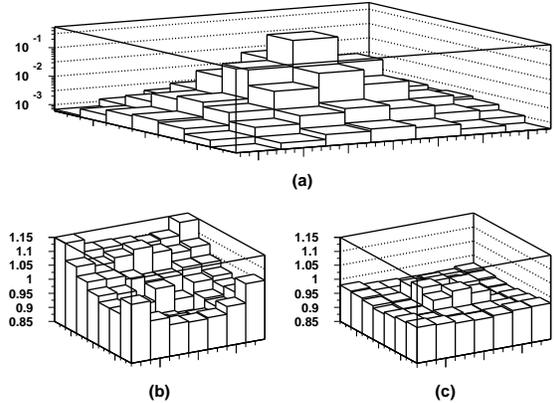}
\caption{Data-MC comparison of fraction of energy in each of the 49 CsI
crystals in an electron shower.  
(a) The fraction of energy in each of the 49 CsI crystals in an electron
shower for data.  (b) 2003 data/MC ratio (c) current data/MC ratio}
\label{fig:ke3shwr2}
\end{center}
\end{figure}

The current Monte Carlo produces a significantly better 
simulation of electromagnetic showers in the CsI.  
Figure \ref{fig:ke3shwr2}
shows the data-MC comparison of the fraction of energy 
in each of the 49 CsI crystals in a
shower relative to the total reconstructed shower energy for 
electrons from $\ke3$ decays.  The majority of the energy is deposited in the
central crystal since the Moliere radius of CsI is 3.8 cm.  
These particular plots are made for 16-32 GeV electrons with incident angles 
of 20-30 mrad, but the quality of 
agreement is similar for other energies and angles.  The data-MC disagreement 
improves from up to 15\% for
the 2003 MC to less than 5\% for the current MC.  This improvement in the
modeling of electromagnetic shower shapes leads to important reductions 
in the systematic
uncertainties associated with the reconstruction of photon showers from
$\kneut$ decays.

\section{Data Analysis}
\begin{figure}
\begin{center}
\includegraphics[width=80mm]{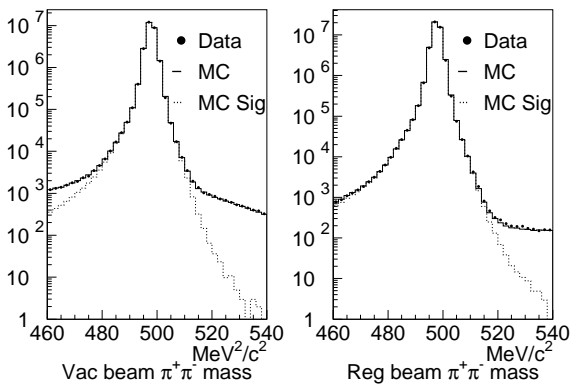}
\end{center}
\caption{\label{fig:massch}
$\pi^+\pi^-$ invariant mass distribution for $\kchrg$ candidate
events. The data distribution is shown as dots, the $\kchrg(\gamma)$ 
signal MC (MC Sig) is shown as dotted histogram and the sum
of signal and background MC is shown as a solid histogram.}
\end{figure}

The $\kchrg$ analysis consists primarily of the reconstruction of tracks 
in the spectrometer.  The vertices and momenta of the tracks are used to 
calculate kinematic quantities describing the decay.  The $\kchrg$
invariant mass distributions for each beam are shown in Figure 
\ref{fig:massch}.

\begin{figure}
\begin{center}
\includegraphics[width=80mm]{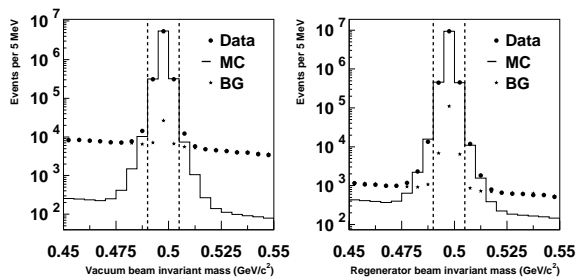}
\caption{$\kneut$ m$_{\pzpz}$ distributions for for data (dots) and signal 
MC (histogram) in the vacuum (left) and regenerator (right) beams.  The sum 
of the background MC is also shown (stars). The dashed lines indicate our 
cuts.}
\label{fig:mkcut}
\end{center}
\end{figure}

To reconstruct 
$\kneut$ decays, we measure the energies and 
positions of each cluster of energy in the CsI.  
A number of 
corrections are then made to the measured particle energies based on our 
knowledge of the CsI performance and the reconstruction algorithm.  The
precision of the CsI energy and position reconstruction is crucial to the
$\kneut$ analysis and has been improved significantly since the previous
publication.  
We use the cluster positions and energies along with the known pion mass 
to determine which pair of photons is associated with which neutral pion 
from the kaon decay and calculate the decay vertex, the center of
energy, and the $\pzpz$ invariant mass.  
The $\kneut$ invariant mass distributions
for each beam are shown in Figure \ref{fig:mkcut}.

\begin{figure}
\begin{center}
\includegraphics[width=80mm]{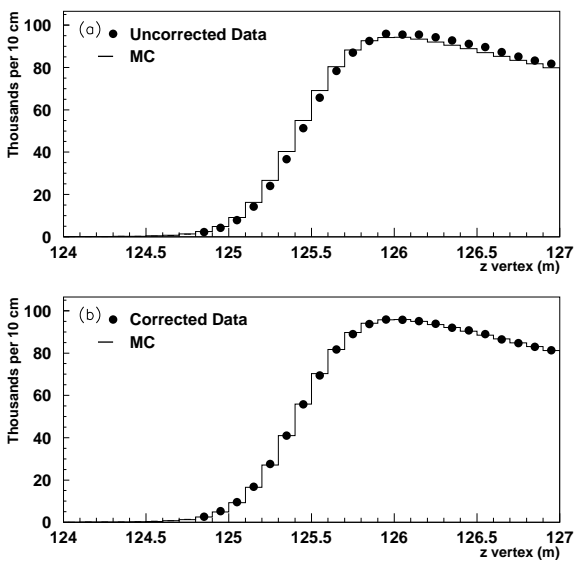}
\caption{Regenerator beam $\kneut$ $z$-vertex distribution near the
regenerator for 1999 data and Monte Carlo.  (a) Uncorrected data.  (b) Data
with energy scale correction applied.}
\label{fig:edgematch}
\end{center}
\end{figure}

\begin{figure}
\begin{center}
\includegraphics[width=80mm]{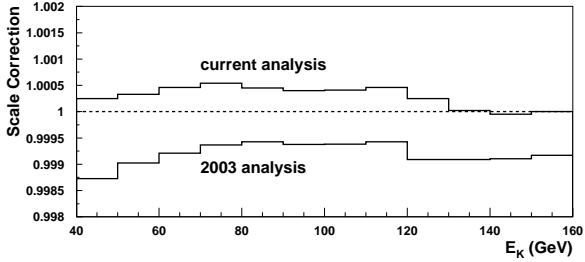}
\caption{Change in the final energy scale adjustment 
relative to the 2003 analysis.  The dashed line represents no data-MC
mismatch.}
\label{fig:scalechange}
\end{center}
\end{figure}

\looseness+1
For $\Kzz$ decays, the $z$ vertex is determined using only the positions
and energies of the four photons in the final state.  Therefore, the 
measured $z$ vertex is dependent upon the absolute energy scale of the 
CsI calorimeter.  The energy scale is set using
electrons from $\ke3$ decays.  A small residual energy scale mismatch between
data and Monte Carlo is removed by adjusting the energy scale in data
such that the sharp edge in the $z$-vertex distribution at the regenerator
matches between data and Monte Carlo as shown in Figure \ref{fig:edgematch}.  
The final energy scale adjustment for 1997 data is
shown as a function of kaon energy in Figure \ref{fig:scalechange};  the
average size of the correction is $\sim$0.04\%.
As a result of improvements to the simulation and reconstruction of clusters, 
the required energy scale adjustment is smaller and less
dependent on kaon energy for low kaon energies than in the previous analysis.

\looseness+1
Background to the $\Kpp$ signal modes is simulated using the Monte Carlo, 
normalized to data outside the signal region, and subtracted.  
In this analysis, we use decays 
from coherently regenerated kaons only; any kaons that scattter with non-zero
angle in the regenerator are treated as background.  Scattering background is
the same for both $\Kpm$ and $\Kzz$ decays so it can be identified 
using the reconstructed transverse momentum of the decay products in charged mode;
we use $\Kpm$ decays to tune the simulation of scattering background on which we
must rely in neutral mode. Non-$\pi\pi$ background
is present due to the misidentification of high branching-ratio decay modes
such as $\ke3$, $\km3$, and $\Kzzz$.    
Background contributes less than 0.1\% of
$\kchrg$ data and about 1\% of $\kneut$ data.



Table \ref{tb:systsummary} summarizes the systematic uncertainties on
$\reepoe$.  We describe the procedure for evaluating several important 
systematic uncertainties below.

\begin{table}[ht]
\centering
\begin{tabular}{l|cc} 
\hline\hline
Source                & \multicolumn{2}{c}{Error on $\reepoe$ ($\eu$)} \\
                      & $\kchrg$ & $\kneut$                             \\ \hline
Trigger               & 0.23     & 0.20                                 \\
CsI cluster reconstruction & --- & 0.75                                 \\
Track reconstruction  & 0.22     & ---                                  \\
Selection efficiency  & 0.23     & 0.34                                 \\
Apertures             & 0.30     & 0.48                                 \\
Acceptance            & 0.57     & 0.48                                 \\
Background           & 0.20     & 1.07                                 \\
MC statistics         & 0.20     & 0.25                                 \\ \hline
Total                 & 0.81     & 1.55                                 \\ \hline
Fitting               & \multicolumn{2}{c}{0.31}                       \\ \hline
Total                 & \multicolumn{2}{c}{1.78}                          \\ \hline\hline
\end{tabular}
\caption{Summary of systematic uncertainties in $\reepoe$.}  
\label{tb:systsummary}
\end{table}

\emph{Acceptance:}
We use the Monte Carlo simulation to estimate
the acceptance of the detector in
momentum and $z$-vertex bins in each beam.  We evaluate the quality of
this simulation by comparing energy-reweighted $z$-vertex distributions
in the vacuum beam between data and Monte Carlo. We fit a line to the 
data-MC ratio of the $z$-vertex distributions and
call the slope of this line, $s$, the acceptance ``z-slope.'' 
A z-slope affects the value of $\reepoe$ by producing a bias between the 
regenerator and vacuum beams because of the very different $z$-vertex 
distributions in the two beams; we use the known difference of the mean
$z$ values for the vacuum and regenerator beams along with the measured
z-slope to evaluate the systematic error on $\reepoe$.

\begin{figure}
\begin{center}
\includegraphics[width=80mm]{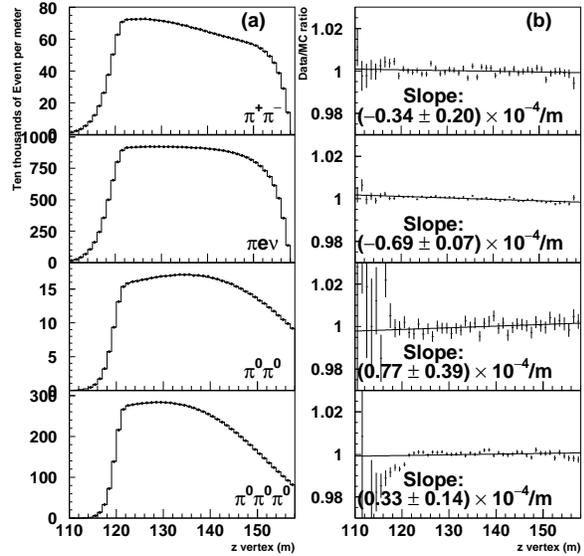}
\caption{Comparison of the vacuum beam $z$ distributions for
data (dots) and MC (histogram).  The data-to-MC ratios on
the right are fit to a line, and the z-slopes (see text) are
shown.  All distributions are for the full data sample used in this
analysis.}
\label{fig:zslopes}
\end{center}
\end{figure}

Figure \ref{fig:zslopes} shows the measured z-slopes for the full $\KLpm$,
$\ke3$, $\KLzz$, and $\Kzzz$ event samples.  We use the $\ppc$ z-slope
to set the systematic uncertainty and measure the $\keth$ z-slope as a
crosscheck.  For neutral mode, we use the high statistics $\zzz$ mode
to set the systematic uncertainty because it has the same type of particles 
in the 
final state as $\pzpz$ and is more sensitive than $\pzpz$ to potential 
problems in the reconstruction due to close clusters, energy leakage at 
the CsI edges, and low photon energies.  

\begin{figure}
\begin{center}
\includegraphics[width=80mm]{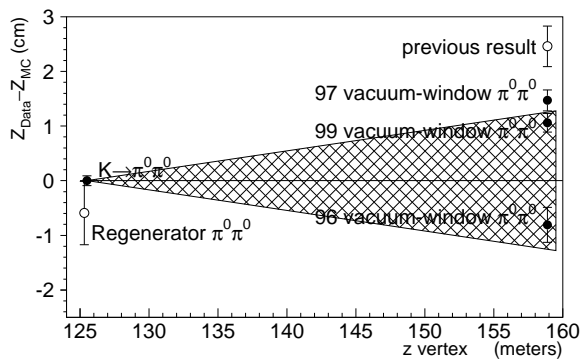}
\caption{Energy scale tests at the regenerator 
and vacuum window.  The difference between the reconstructed $z$ positions 
for data and MC is plotted for the $\kneut$, regenerator $\pzpz$, and vacuum 
window $\pzpz$ samples.  The solid point at the regenerator edge is the 
$\kneut$ sample; there is no difference between data and MC by
construction.  The open point at the regenerator edge is the average shift 
of the regenerator $\pzpz$ samples for all three years.  The points at the 
vacuum window are the shifts for the vacuum window samples for each year 
separately.  The hatched region shows the range of data-MC shifts covered 
by the total systematic uncertainty from the energy scale.
For reference, the data-MC shift at the vacuum window from the 2003
analysis is also plotted.}
\label{fig:escalesyst}
\end{center}
\end{figure}

\emph{Energy Scale:}
The final energy scale adjustment ensures that the energy scale matches 
between data and MC at the regenerator edge, but we must check whether 
the data and MC energy scales remain matched for the full length of the 
decay volume.  
We check the energy scale at the downstream end of the decay region by 
studying the $z$-vertex distribution of $\pzpz$ pairs produced
by hadronic interactions in the vacuum window in data and MC.  
To verify that this type of 
production has a comparable energy scale to $\kneut$, we also study the 
$z$-vertex distribution of hadronic $\pzpz$ pairs produced in the regenerator.
The data-MC comparisons of reconstructed $z$
vertex for these samples are shown in Figure \ref{fig:escalesyst}.

To convert these shifts to an uncertainty in
$\reepoe$, we consider a
linearly varying energy scale distortion such that no adjustment 
is made at the regenerator edge and the $z$ shift at the vacuum window is 
that measured by the hadronic vacuum window sample.  The average energy scale distortion
we apply is shown by the hatched region in Figure \ref{fig:escalesyst}.
We rule out energy scale distortions that vary non-linearly as a function 
of $z$ vertex as they introduce data-MC discrepancies in other distributions.
The systematic error on $\reepoe$ due to uncertainties in the $\kneut$ 
energy scale is 0.65$\eu$; this is a factor of two smaller than in the previous
analysis.

\begin{figure}
\begin{center}
\includegraphics[width=80mm]{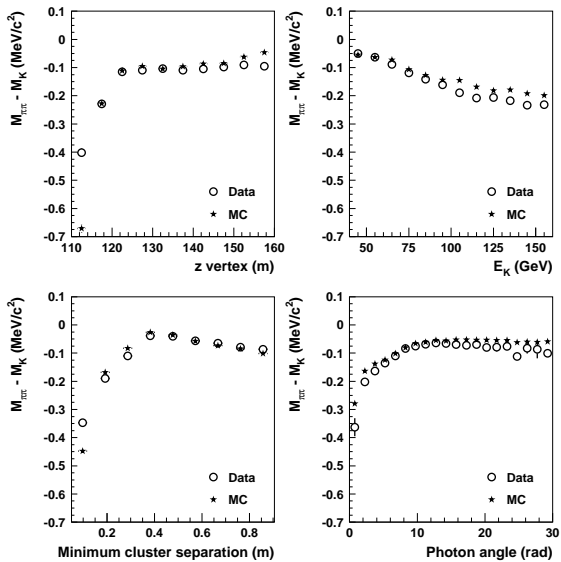}
\caption{Comparisons 
of the reconstructed kaon mass vs $z$-vertex (top left), kaon energy 
(top right), minimum cluster 
separation (bottom left), and photon 
angle (bottom right) for 1999 data and MC.
The values plotted are the difference between the reconstructed
kaon mass for each bin and the nominal PDG kaon mass.}
\label{fig:nonlin1}
\end{center}
\end{figure}

\begin{figure}
\begin{center}
\includegraphics[width=80mm]{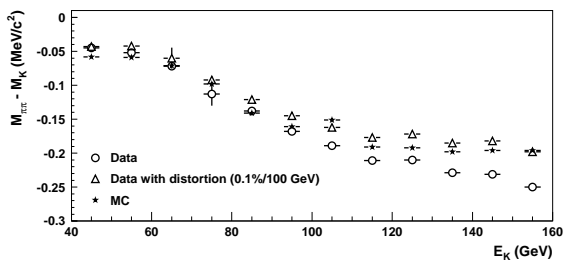}
\caption{Effect of 0.1\%/100 GeV distortion on M$_K$ vs E$_K$ for 1999 data.
The values plotted are the difference between the reconstructed
kaon mass for each bin and the nominal PDG kaon mass.}
\label{fig:nonlin2}
\end{center}
\end{figure}

\emph{Energy Non-linearity:}
Some reconstructed quantities in the analysis do not depend on the CsI
energy scale, but are sensitive to energy non-linearities.
To evaluate the effect of energy non-linearities on the reconstruction,
we study the way the reconstructed kaon mass varies with reconstructed 
kaon energy, 
kaon $z$ vertex, minimum cluster separation, and incident photon angle.  
Data-MC comparisons for these distributions for the 1999 data sample 
are shown in Figure 
\ref{fig:nonlin1}.  To measure any bias resulting from the nonlinearities
that cause the small data-MC
differences seen in these distributions, we investigate adjustments to the 
cluster energies 
that improve
the agreement between data and MC in the plot of reconstructed kaon mass 
vs kaon energy.  We find that a 0.1\%/100 GeV distortion produces the best 
data-MC 
agreement for the 1997 and 1999 datasets.  
Figure \ref{fig:nonlin2} shows the 
improvement in data-MC agreement with this distortion applied 
to 1999 data.
The data-MC agreement in the reconstructed kaon mass as a function of
kaon energy has been significantly improved compared to the previous analysis 
in which a 0.7\%/100 GeV distortion was required for 1997 data.

\section{Results}
The final KTeV measurement of $\reepoe$ for the full 1996, 1997,
and 1999 combined dataset is:
\bqa
\reepoe & = & [19.2 \pm 1.1(stat) \pm 1.8(syst)]\eu \\ \nonumber
        & = & [19.2 \pm 2.1]\eu.                     \\
\eqa

\begin{figure}
\begin{center}
\includegraphics[width=80mm]{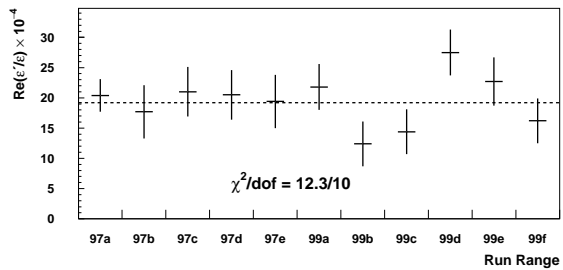}
\caption{$\reepoe$ in subsets of the data sample.  Each point is
statistically independent.  The dashed line indicates the value of $\reepoe$ for the full data sample.  The 97a 
run range includes the 1996 $\kneut$ data.}
\label{fig:eperuns}
\end{center}
\end{figure}

\begin{figure}
\begin{center}
\includegraphics[width=80mm]{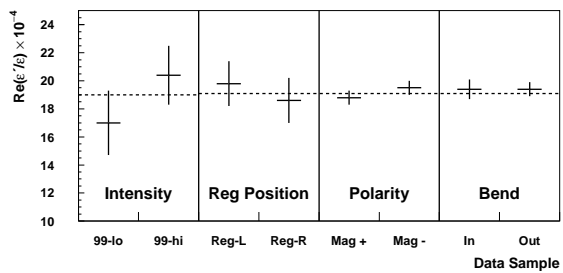}
\caption{$\reepoe$ consistency with beam intensity, regenerator position, 
magnet polarity, and track bend.   The low and high intensity samples are from 1999
only and have
average rates  of $\sim1\times10^{11}$ protons/s and  $\sim1.6\times10^{11}$ 
protons/s, respectively.  
Reg-left and reg-right refer to the position of the regenerator beam in 
the detector.  Mag+ and Mag- are
the magnet polarity and in/out are the bend of the tracks in the magnet.  
In each of these subsets the $\kneut$ sample is common to both fits; 
the errors are estimated by taking the quadrature difference with
the error for the full dataset.  The dashed lines
indicate the value of $\reepoe$ in the appropriate full data sample.}
\label{fig:epechecks}
\end{center}
\end{figure}

\begin{figure}
\begin{center}
\includegraphics[width=80mm]{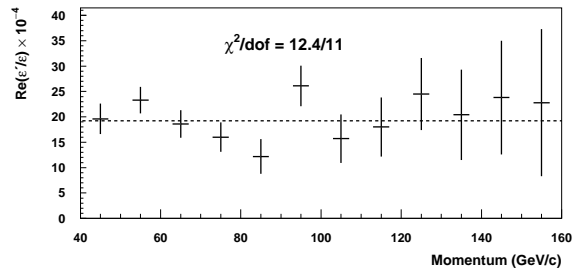}
\caption{$\reepoe$ in 10 GeV/c momentum bins.
The dashed line indicates the value for the full data sample.}
\label{fig:epepbins}
\end{center}
\end{figure}

We perform several checks of our result by breaking the data 
into subsets and checking the
consistency of the $\reepoe$ result.  To check for 
any time dependence, we break the data into 11 run ranges with roughly equal 
statistics.  We divide the data in half based on beam intensity, regenerator
position, magnet polarity, and direction in which the tracks bend in the
magnet.  We check for dependence of the result on kaon momentum by breaking 
the data into twelve 10 GeV/c momentum bins.
The $\reepoe$ 
results for these tests are shown in Figures \ref{fig:eperuns}, 
\ref{fig:epechecks}, and \ref{fig:epepbins}.  We find 
consistent results in all of these subsamples.

We also measure the kaon parameters $\tauS$, $\delm$,
$\phiep$, $\reepoe$, and $\imepoe$ in a single, z-binned fit.  The systematic
uncertainties are evaluated using a procedure identical to that used
for the $\reepoe$ measurement.  CPT invariance is
imposed {\it a posteriori} including the total errors of the parameters 
with their correlations to obtain a precise measurement of 
$\delm$ and $\tau_S$.  The results are:
\begin{equation}\label{eq:zresults}
\begin{array}{lcl}
  \delm\,|_{\rm cpt}  &=& [\dmswval \pm \dmswerr ] \times 10^{-12}~{\rm s}, \\
  \tauS\,|_{\rm cpt}  &=& [\tsswval \pm \tsswerr ] \times 10^6 {\rm \hbar/s}, \\
  \phipm &=& [\phpmval \pm \phpmerr]\degs,\\
  \phizz &=& [\phzzval \pm \phzzerr]\degs, \\
  \delta \phi &=& \phiep - \phi_{SW} = [\dphswval \pm \dphswerr]\degs, \\
  \Delta \phi &=& - 3 \imepoe = [\dphcptval \pm \dphcpterr ]\degs.\\
\end{array}
\end{equation}

\begin{acknowledgments}
We gratefully acknowledge the support and effort of the Fermilab staff
and the technical staffs of the participating institutions.  This work
was supported in part by the U.S. Department of Energy, The National
Science Foundation, and the Ministry of Education and Science in Japan.

\end{acknowledgments}

\bigskip 

\begin{thebibliography}{9}   
\bibitem{ochs} W. Ochs, $\pi N$ Newsletter 3, 25 (1991).
\bibitem{prl:731} L.K. Gibbons et al. (E731), Phys. Rev. Lett. 70, 1203 (1993).
\bibitem{pl:na31} G.D. Barr et al. (NA31), Phys. Lett. B317, 223 (1993).
\bibitem{prl:pss} A. Alavi-Harati et al. (KTeV), Phys. Rev. Lett. 83, 22 (1999).
\bibitem{na48:reepoe} A. Lai et al. (NA48), Eur Phys. J. C 22, 231 (2001).
\bibitem{prd03} A. Alavi-Harati et al. (KTeV), Phys. Rev. D67, 012005 (2003).
\end{thebibliography}

\end{document}